\documentclass[preprint]{emulateapj}

\slugcomment{Submitted to the Astrophysical Journal}
\shortauthors{Zaritsky et al.}
\shorttitle{}

\begin{document}
\title{Evidence for Two Distinct Stellar Initial Mass Functions : \break Revisiting the Effects of Cluster Dynamical Evolution}
  
\author{Dennis Zaritsky}
\affil{Steward Observatory, University of Arizona, 933 North Cherry Avenue, Tucson, AZ 85721}

\author{Janet E. Colucci}
\affil{Department of Astronomy and Astrophysics, 1156 High Street, UCO/Lick Observatory, University of California, Santa Cruz, CA 95064}

\author{Peter M. Pessev}
\affil{Gemini South Observatory, c/o AURA Inc., Casilla 603, La Serena, Chile}

\author{Rebecca A. Bernstein}
\affil{Department of Astronomy and Astrophysics, 1156 High Street, UCO/Lick Observatory, University of California, Santa Cruz, CA 95064}

\author{Rupali Chandar}
\affil{Department of Physics and Astronomy, The University of Toledo, 2801 West Bancroft Street, Toledo, OH, 43606}

\email{dzaritsky@as.arizona.edu}


\begin{abstract} 
We measure the velocity dispersions of six, galactic globular clusters using spatially integrated spectra, to test for the effects of internal dynamical evolution in the stellar mass-to-light ratios, $\Upsilon_*$, of star clusters. In particular, we revisit whether the low values of $\Upsilon_*$ we found in our previous study, from which we concluded that there are at least two population of stellar clusters with distinct stellar initial mass functions, are artificially depressed by relaxation driven mass loss. The combination of our previous sample of five old clusters and these six now provides an order of magnitude range in cluster mass with which to explore this issue. We find no relationship between cluster mass, or relaxation time, and $\Upsilon_*$. Because relaxation is mass dependent, we conclude that the values of $\Upsilon_*$ for these clusters are not strongly affected by dynamical effects, and so confirm the presence of the population of clusters with low $\Upsilon_*$. 
\end{abstract}

\keywords{stars: formation, luminosity function, mass function; galaxies: fundamental parameters, evolution}

\section{Introduction}
\label{sec:intro}

A variety of independent lines of evidence now point to a stellar initial mass function (IMF) that can vary from one environment to the next. In elliptical galaxies the study of population-dependent spectral features \citep{vandokkum,spiniello} and detailed dynamical modeling \citep{cappellari12} suggest varying IMFs that are different than those measured in the local Galactic disk \citep{bastian}, in the Small Magellanic Cloud deep star counts suggest an IMF that does not turn over at low masses \citep{kalirai}, and in the study of stellar clusters anomalies are found both among M31 clusters \citep{strader} and in those of the Milky Way and its satellites \citep[][hereafter Paper I]{z12}. While it is necessary to continue to explore the IMF in all of these environments and provide more evidence in support of these initial findings, the study of clusters presents one significant advantage over that in the other environments. Because star clusters are the most likely to contain only a simple stellar population, of a single age and metallicity, these systems provide the possibility of a relatively direct approach to unraveling the cause of IMF variations --- if the variations are confirmed. 

In Paper I we found a bifurcation in the apparent properties of star clusters. Specifically, we found that younger clusters (age $< 10$ Gyr) typically, but not exclusively, have discretely larger values of the stellar mass-to-light ratio, $\Upsilon_*$, for their age than do the older clusters. The explanation we proposed is that there exist (at least) two populations of clusters that have different IMFs. There are, of course, a number of alternate explanation for the difference in properties, which we discussed in Paper I, but here we present new data and an empirical exploration of one of the principal potential sources of systematic error, the impact of internal dynamical evolution on $\Upsilon_*$.

Older clusters, which have suffered the greatest degree of two-body relaxation, may have preferentially lost large numbers of low-mass stars, and thereby have lower $\Upsilon_*$ than models of stellar evolution would predict for a given IMF. As such, one wonders if the discordant properties of the  younger and older clusters can be reconciled. In Paper I we explored this possibility quantitatively using published models of stellar cluster dynamical evolution \citep{anders}, but the application of those models has large uncertainties because certain input parameters are not well constrained. Here we present observations of six additional old clusters, which more than doubles our sample of older clusters observed and analyzed in the exact same manner. Furthermore, these new clusters now enable us to probe the effects of dynamical evolution over a larger range of cluster mass. Because the effects of two-body relaxation are most pronounced in lower mass clusters, a sample with a large mass range provides an empirical test of the effects of two-body relaxation.  We present the data in \S2, discuss the derived masses, $\Upsilon_*$ and examine the dependence of $\Upsilon_*$ on mass and relaxation time in \S3. We find no dependence, demonstrating that two-body relaxation has had a minimal impact on our results and that the values derived for $\Upsilon_*$ for our old clusters are not artificially deflated. As such, we confirm the existence of the lower $\Upsilon_*$ population of clusters. We have therefore addressed one of the most significant source of systematic uncertainty in our previous results. Future work will present more data related to the population of stellar clusters with large values of $\Upsilon_*$.
   
\section{The Data}

Our spectroscopic data come from a set of observations taken with the Las Campanas du Pont telescope (100-inch), primarily for the purpose of chemical abundance studies \citep{colucci1,colucci2}. 
We use the compilation of \cite{mclaughlin} for the necessary ancillary data (age, half light radius, luminosity, and metallicity). When model data are used from the compilation, we choose results obtained using the Wilson models \citep{wilson}, which \cite{mclaughlin} demonstrate are superior in fitting the radial surface brightness profiles of these clusters. All of these choices follow our procedure in Paper I and, as then, our conclusions are insensitive to 
the choice of the Wilson model fits.

We obtained our spectra with the echelle spectrograph at the 100-inch at Las Campanas (du Pont) during dark time in 2000 and 2001. To obtain integrated spectra of the clusters, we utilize the same spectroscopic drift scan technique described by \cite{colucci1}. To summarize, we set the telescope in motion to raster scan the slit across the cluster during the exposure, defining both the angular length and height of the raster (both set to the same number). The exposure time then sets the rate of the scan, such that the full scan is completed within the allotted exposure time. 
The echelle slit is 1\arcsec x 4\arcsec,  allowing uniform coverage of a 32 $\times$ 32 arcsec$^2$ high-surface brightness region of the cluster. We took multiple exposures to homogenize the scanned cluster region and for cosmic ray removal.
The spectra have a wavelength coverage of approximately 3700 to 7800 \AA, with declining sensitivity and spectral resolution toward the blue end.  We reduced the spectra using standard IRAF routines (see Colucci et al. 2011), including the scattered light subtraction described in detail in \cite{mcwilliam}. Further details of the observations and data reduction can be found in \cite{colucci1}. 

The spectra are of the same high quality as in our original study (Paper I) because the total exposure times are also typically around 10,000 seconds (see Table \ref{tab:clusters}). Examples of similar spectra and the spectral fitting described below are given in the original paper. There are no substantive differences between those data and that described here. This similarity is confirmed by the internal velocity uncertainties derived from the fitting, described below, that are quantitatively comparable to those obtained for the clusters in Paper I. The S/N of the spectra varies among the objects and as a function of wavelength but generally exceeds 20.

\begin{deluxetable*}{lllrrrrrrrr}
\tablewidth{0pt}
\tablecaption{Stellar Cluster Data}
\tablehead{
\colhead{NGC} &
\colhead{Host} &
\colhead{Tel.} &
\colhead{$t_{exp}$} &
\colhead{log(age)} &
\colhead{log(L$_{V}$)}&
\colhead{$\langle Fe/H \rangle$} &
\colhead{$r_h$} &
\colhead{log(I$_h$)} &
\colhead{$\sigma$} &
\colhead{$\Upsilon_{*}$} \\
&&&[s]&[Gyr]&[L$_{\odot}$]&&[pc]&$[L_\odot pc^{-2}]$&km s$^{-1}$&$[\odot]$\\
}
\startdata
0104 & MW &  C100 & 10800  &  10.11 & 5.66  &$-$0.76 & 3.00  & 3.91 & $11.47^{+0.19}_{-0.21}$&  0.96$^{+0.03}_{-0.03}$ \\
0362 & MW &  C100 & 10800 &  10.11 & 5.20  &$-$1.16 & 1.55  & 4.02 & $ 9.15^{+0.35}_{-0.41}$&  0.92$^{+0.07}_{-0.08}$ \\
2808 & MW &  C100 & 10440 &  10.11 & 5.63 &$-$1.16 & 1.98  & 4.24 & $12.96^{+0.50}_{-0.51}$&  0.87$^{+0.07}_{-0.07}$  \\
6093 & MW &  C100 &  7200 &  10.11 & 5.16 &$-$1.75 & 1.57  & 3.97 & $9.49^{+0.46}_{-0.51}$&1.09$^{+0.11}_{-0.11}$\\
6388 & MW &  C100 &  10800 & 10.11 & 5.64 &$-$0.60 & 1.41 & 4.54 & $17.99^{+0.46}_{-0.52}$&  1.18$^{+0.06}_{-0.07}$ \\
6752 & MW &  C100 &  10800  & 10.11 & 5.02 &$-$1.56 & 2.22 & 3.53 & $6.62^{+0.34}_{-0.40}$&  1.03$^{+0.11}_{-0.12}$ \\
\enddata
\label{tab:clusters}
\end{deluxetable*}

\subsection{Measuring Velocity Dispersions}

A discussion motivating our approach is presented in Paper I, and we briefly review it here.

We fit a Gaussian broadening function to each of the same set of lines selected in Paper I. 
We reject any lines in the object spectra that are either clearly blends or suffer some other complication and, after fitting, we reject lines that do not produce an acceptable fit, where acceptable is defined by $\chi_\nu^2 < 2.3$. To calculate $\chi_\nu^2$ we adopt a per pixel uncertainty determined from the fluctuations about a flat continuum in line-free areas of the spectrum. However, we adopt the same uncertainty value for the full spectral range for any given cluster. Our results are not highly sensitive to this value because we only use these $\chi_\nu^2$ values to remove questionable fits from further consideration, and visual inspection confirms that those lines that have been rejected by this criteria are poorly fit.

Once the line fitting is complete, we calculate the overall best value of $\sigma$ using the same approach as in Paper I, where we 
place different weights on downward and upward uncertainties because larger apparent broadenings can be caused by various systematic issues (blended lines, poor continuum fit, focus errors). 
The final value of $\sigma$ for each cluster is calculated using the average of all the measurements 
and $\chi^2/N$, where $N$ is the number of data points, with uncertainties in any individual measurement defined as resulting in $\chi^2/N = 1$. To down weight measurements that are inflated by blends or focus errors, we set upward uncertainties to be 3 times larger than downward ones. 
The uncertainty on our final ``mean" $\sigma$ is  derived by identifying the range of $\sigma$'s that generate $\chi_\nu^2 - \chi^2_{nu,min} < 2.71$, corresponding to the 90\% confidence interval (see Table \ref{tab:clusters}). For a more complete description and discussion of the issues involved we refer the reader to Paper I.

\begin{figure}[htbp]
\plotone{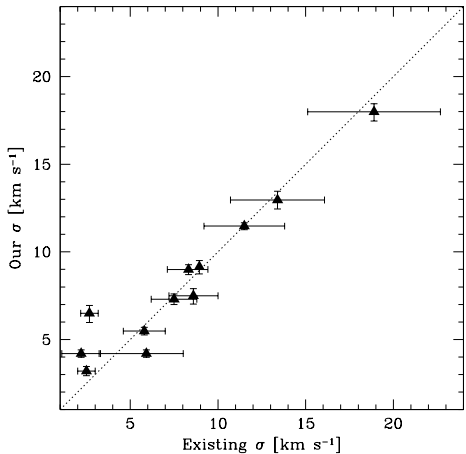}
\caption{A comparison of our measurement of the cluster velocity dispersion, $\sigma$, to previously published measurements by independent investigators. The comparison includes clusters of all ages (from Paper I and this study). The line represents the 1:1 relation. The agreement is excellent, but our new values provide the improved precision necessary to uncover subtle differences in $\Upsilon_*$.}
\label{fig:sigcomp}
\end{figure}

For the subset of our clusters that have been observed previously, we compare our $\sigma$ measurements to pre-existing values. 
We obtain previous measurements from the compilation of \cite{mclaughlin}, which includes literature values of $\sigma$ for 9 of our clusters (over the entire age range), and other sources that include 3 more \citep{meylan,fischer,lane}. For NGC 362, a value of the dispersion was not presented by \cite{fischer} so we calculate the standard deviation of the individual stellar velocities they provide, after excluding the sources they identified as either non-members or binary stars.
The external comparison is positive in that 9 of 12 (75\%) of the measurements agree to 1$\sigma$ and 
11 of 12 (92\%) agree to 2$\sigma$ (Figure \ref{fig:sigcomp}). As we discussed in Paper I,  the published dispersion of the one significant outlier, NGC 1866, is highly sensitive to the inclusion or exclusion of a single star \citep{lane}.

\section{Determining Masses}

In Paper I we applied the prescription presented and tested by \cite{walker} to derive masses for spheroidal galaxies. As discussed previously in more detail, our finding that their prescription, when expressed in the formalism of galaxy scaling relations,
\begin{equation}
\log r_h = 2\log \sigma - \log I_h - \log \Upsilon_h - 0.73,
\label{eq:walker}
\end{equation}
where $r_h$, $I_h$, and $\Upsilon_h$ are the half light radius, the surface brightness within that radius, and the mass-to-light ration within that radius, almost exactly matches the empirical finding of a scaling relation that applies to all stellar systems from stellar clusters to massive ellipticals
\begin{equation}
\log r_h = 2\log \sigma - \log I_h - \log \Upsilon_h - 0.75.
\label{eq:fm}
\end{equation}
\citep{z08,z11} provides additional confidence in this method. 
The empirical results verify that there is little scatter ($\sim 0.1$ dex) about this relationship for objects ranging from globular clusters to massive elliptical galaxies. We also demonstrated in Paper I that this method results in values of $\Upsilon_h$ in good agreement with published values calculated using more detailed dynamical models.
Using Equation \ref{eq:fm}, we evaluate $\Upsilon_h$ for the six clusters of this study and present the results in Table \ref{tab:clusters}. 
For systems without dark matter, which we presume includes these clusters, $\Upsilon_* \equiv \Upsilon_h$. All photometric quantities are presented for the $V$ band.
The uncertainties in $\Upsilon_*$ are calculated using only the uncertainty in $\sigma$, as the internal uncertainties on the other parameters are proportionally much smaller. 

\subsection{Dynamical Evolution of Clusters and Its Impact on $\Upsilon_*$}
\label{sec:dynamical}

A range of dynamical process affect stellar clusters. Of particular relevance to the type of study we are conducting is the evaporation of low-mass stars via two-body interactions, which will depress the value of $\Upsilon_*$ and has the strongest impact on the oldest clusters \citep{spitzer, kruijssen08}. In Paper I, we used the models of 
\cite{anders} who calculated the disruption time for clusters, $t_d$, which depend on the cluster mass and local (external) tidal field. Applying the models, however, depends on the choice of the normalization factor $t_4$, which is the age at which a 10$^4$ M$_\odot$ cluster loses 95\% of its mass under the corresponding conditions.
Although each cluster should be assigned its own value of
$t_4$, corresponding on its internal density, orbit, and tidal field strength, we had to adopt general values of $t_4$ without much guidance. Nevertheless, we proceeded to calculate the correction factors for $\Upsilon_*$, but found that in general these were small and even in the most extreme cases they were less than a factor of two, which is insufficient to reconcile the differences between the two populations of clusters. However, because of the uncertainty in the selection of $t_4$ and the dependence on idealized models, an unsatisfactory level of uncertainty remains.

The data described here, in combination with that in Paper I, provides an opportunity to reexamine the question in a more empirical manner. The modeled relaxation effects are inversely proportional to the mass of the cluster, so we expect low mass clusters to have suffered the most. Specifically,  in the \cite{anders} models $t_d \propto M^{0.62}$.  With the new data presented here, the sample of old clusters we have has more than doubled and spans an order of magnitude in mass, suggesting disruption times that should differ by a factor of 4. Therefore, if some of our oldest clusters, age $\sim$ 13 Gyr, have lived for a significant fraction of their disruption time, then the most massive in the sample have disruption times that are $> 50$ Gyr, and so should show negligible effects of relaxation. We therefore expect that if cluster values of $\Upsilon_*$ are affected by relaxation, the effect should be significantly larger at the low mass end of our range.

\begin{figure}[htbp]
\plotone{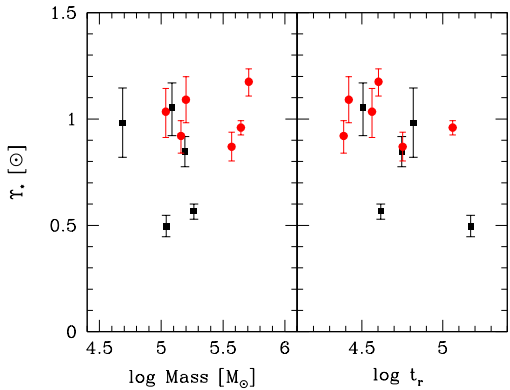}
\caption{$\Upsilon_*$ vs. cluster mass and relaxation time. Relaxation time is presented in arbitrary units. We find no discernible trend between $\Upsilon_*$ and cluster mass or relaxation time. The lack of trends argues against the importance of two-body relaxation effects because in such a scenario $\Upsilon_*$ should be lower in the low mass and/or short relaxation time systems. Clusters from Paper I are plotted as black squares, while the clusters presented here are plotted as red circles.}
\label{fig:massdep}
\end{figure} 

 In Figure \ref{fig:massdep}, we plot the relationship between $\Upsilon_*$ and cluster mass. We find no evident mass dependence. This result does not rule out dynamical evolution in all systems, for example the two clusters with lower than average $\Upsilon_*$ may have experienced some. The variations in $\Upsilon_*$ seen among the older clusters in other studies, and presumably in ours, can be explained by two-body relaxation effects \citep{leigh}, and our results do not contradict this claim. Our conclusion is that dynamical evolution cannot result in the much larger effect required to reconcile these observed values of $\Upsilon_*$ with those of our other cluster population (Paper I). The only possible way to avoid this conclusion is to evoke a detailed inverse correspondence between the strength of tidal effects and the mass of the clusters (with tidal effects being stronger on the the more massive clusters and canceling out the mass dependence of the two-body relaxation effects).
We see such a correspondence as contrived, particularly because some of the older clusters (from Paper I) are in the Large Magellanic Cloud, which should provide a significantly different tidal environment than for those clusters in the Milky Way. 

In Figure \ref{fig:massdep}, we also plot $\Upsilon_*$ against a more direct estimate of the relaxation time, $t_r \propto Nt_{cross}$, where $N$ is the number of stars and is proportional to $M_*$ and $t_{cross} \propto r_h/\sigma$. Again, we find no dependence that leads us to conclude that our previous estimates of $\Upsilon_*$ were biased low by dynamical evolution.

\begin{figure}[htbp]
\begin{center}
\plotone{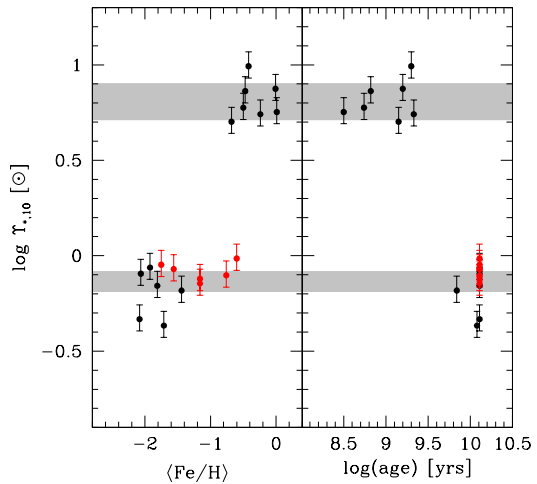}
\caption{Evidence for two initial mass functions revisited. We plot  $\Upsilon_{*,10}$, the value of $\Upsilon_*$ at 10 Gyr for each cluster, obtained as described in Paper I, versus iron abundance and age. In red we highlight the clusters presented in this paper.  The shaded regions represent the standard deviations from the mean for that corresponding population. These are different than shown in Paper I because they have been recalculated from the data presented in this Figure.}
\label{fig:m2l} 
\end{center}
\end{figure}

Finally, in Figure \ref{fig:m2l} we reprise Figure 10 from Paper I. The plot shows the value of $\Upsilon_*$ that a cluster will have (or had) at an age of 10 Gyr, $\Upsilon_{*,10}$. We describe this calculation in Paper I, although for the clusters presented here this correction is minor because they are all only slightly older than 10 Gyr. We have made several alterations from the version of the plot shown in Paper I. First, we include now only clusters for which we have measured the velocity dispersions (previously we had augmented our set with velocity dispersions presented in \cite{mclaughlin}). This choice provides a degree of homogeneity to the data that facilitate comparisons among clusters. Second, we have removed the correction for dynamical evolution because we now consider it to be minor, for these clusters,  and highly uncertain. Third, we recalculate the uncertainty regions for the lower $\Upsilon_{*,10}$ population including these new data and excluding the previously included literature data. 

The distinction between the two populations continues to be clear. As noted in Paper I based on literature data, but now confirmed with our data, there is overlap among the populations along the metallicity axis. Metallicity alone, therefore, cannot be the cause for different IMFs. Additional data  that we are now collecting and processing will attempt to expand the sample at intermediate ages, which will help address the question of whether there is an analogous overlap of populations along the age axis.

We close now by evaluating the gains obtained by our re-observations of clusters that already have velocity dispersion measurements in the literature. The original reason for observing those clusters was to establish that there is no systematic difference in the velocity dispersions obtained through our integrated light measurements and those in the literature based on individual stars. Figure \ref{fig:sigcomp} confirms that there is no difference. Further examination of that Figure also highlights that we are claiming significantly improved precision in our measurement of the velocity dispersions. This improvement, if it can be confirmed, is critical because $\Upsilon$ has a $\sigma^2$ dependence and because among the literature measurements of the required parameters $\sigma$ has the largest fractional uncertainty. As such, the uncertainty in $\sigma$ dominates the uncertainty in $\Upsilon$. Without independent measurements of $\sigma$ that are of comparable or better precision, we cannot directly confirm our uncertainty estimates. However, indirectly we find that the scatter in the calculated $\Upsilon_{*,10}$ values for the ancient clusters obtained with literature $\sigma$'s (Figure 10, Paper I) is nearly 5 times larger than that for our sample (Figure \ref{fig:m2l}), suggesting that the dispersion in our $\sigma$ measurements is indeed significantly smaller that that of previous measurements and that further observations of clusters along the lines presented here, even if they have existing literature values of $\sigma$, is of value.

\section{Summary}

We have measured the velocity dispersions of six, galactic globular clusters using spatially integrated spectra, to test for the effects of internal dynamical evolution in the stellar mass-to-light ratios, $\Upsilon_*$, of star clusters. We conclude, based on the lack of any detectable variation in $\Upsilon_*$ with cluster mass, that dynamical relaxation is not affecting the entire population of clusters. This finding addresses one of the principal potential sources of systematic uncertainty in our previous argument for distinctly different IMFs among two sets of stellar clusters in the Milky Way galaxy and its satellites \citep{z12}. 
Because additional concerns remain, and the possibility of a non-universal IMF is sufficiently important, we are obtaining more data on other clusters to further test our original claims. In the meantime, evidence for variations of the IMF is appearing in a number of environments that is coming from a wide range of independent observations \citep{vandokkum,strader,cappellari12,kalirai}. We advocate increased caution when interpreting extragalactic observations using a single IMF.

\begin{acknowledgments}

DZ acknowledges financial support from 
NSF grant AST-0907771 and NASA ADAP NNX12AE27G. R. C. acknowledges support from NSF through CAREER award 0847467.

\end{acknowledgments}

\end{document}